\documentclass{PoS}
\usepackage{amsmath}
\usepackage{wrapfig}
\usepackage{subfig}

\title{Testing volume independence of large N gauge theories on the
lattice}

\ShortTitle{Testing volume independence of large N gauge theories on
the lattice}

\author{\speaker{Antonio Gonz\'alez-Arroyo}$^{ab}$ \\
\llap{$^a$}Instituto de F\'{\i}sica Te\'orica UAM/CSIC, C/ Nicol\'as Cabrera 13-15\\
       Universidad Aut\'onoma de Madrid, E-28048--Madrid, Spain \\
\llap{$^b$}Departamento de F\'{\i}sica Te\'orica, C-15 \\
       Universidad Aut\'onoma de Madrid, E-28049--Madrid, Spain\\
E-mail: \email{antonio.gonzalez-arroyo@uam.es}}       
\author{Masanori Okawa$^c$ \\      
\llap{$^c$}Graduate School of Science, Hiroshima University\\
Higashi-Hiroshima, Hiroshima 739-8526, Japan\\
E-mail: \email{okawa@sci.hiroshima-u.ac.jp}}

\abstract{For a pure  SU(N) gauge theory on the lattice we test 
if the expectation values of small Wilson loops become volume
independent in the large N limit.
}

\FullConference{The 32nd International Symposium on Lattice Field Theory,\\
		23-28 June, 2014\\
		Columbia University New York, NY}

\begin{document}

\section{The Question}
%
Are gauge theories in the large $N$ limit volume independent?
Or put in different words: Are finite volume corrections subleading 
in the large $N$ limit? These questions were advanced many years 
ago by Eguchi and Kawai~\cite{EK} (EK), when they realized that, under 
certain hypothesis, the loop equations did not depend on the size of
the box. Furthermore, the arguments were valid for the lattice model 
itself. No doubt that if the statement is true at all values of the 
lattice coupling, the result should also hold in the continuum limit.

For years the previous question has been debated and different ways
have been put forward to try to realize this idea. A good deal of
effort has been put into understanding whether the specific conditions 
on which Eguchi and Kawai proof is based were met. Most importantly 
there is the question of center symmetry: invariance under a Z(N)
transformation of the Polyakov loops in each direction. The symmetry
guarantees that the expectation values of (non-zero winding) 
Polyakov loops vanish, so that these observables act as order parameters.  

Recently~\cite{GAO} we have analysed the problem in a direct way: 
computing lattice observables for different sizes and values of $N$
and testing if the size dependence goes to zero at large $N$. The
advantage of this idea is that we test volume independence directly 
and not indirectly via the conditions of EK proof. Hence, it
allows to measure the size of the corrections, namely the finite volume 
dependence at finite $N$. From a practical viewpoint, this is basic  
information one needs when  making use of the volume independence
property. In this  talk we will present a brief summary of our results.

\section{The models}
In Ref.~\cite{EK} the simplest  model with Wilson action 
in a periodic hypercubic lattice was adopted. Volume independence 
meant that the single-site model was equivalent to the infinite volume
one at large N. This is certainly false as pointed out in
Ref.~\cite{bhanot}. The failure lies at weak coupling where perturbative 
methods are good approximations. The classical vacua of the model
(generically) break the $Z^4(N)$ symmetry spontaneously. In a trivial 
fashion one can verify that the plaquette expectation value, for example,
has a finite volume correction which does not go to zero at large $N$.

Several ideas have been put forward over the years to find valid
implementations of the  volume independence idea. One of the earliest 
proposals, introduced  by the present authors,  was based on a very
simple modification of the original proposal~\cite{TEK1,TEK2}. 
The point is to  use twisted  boundary conditions instead of periodic
ones which is perfectly compatible with  EK proof. Furthermore, 
it does not add  any computational or  fundamental complication to 
the EK proposal. With suitable choices of the twist tensor one can 
verify that at weak coupling centre symmetry is broken down to a
subgroup large enough to preserve the validity of volume reduction at
large $N$. In particular, if one opts for preserving as much as  possible
the isotropy among the directions of space-time, one should take
$N=\hat{L}^2$ where $\hat{L}$ is an integer. Then the classical minima
respect $Z^4(\hat{L})$ symmetry which ensures that all Polyakov loops
with windings that are not multiples of $\hat{L}$ should vanish.  This 
choice of twist called {\em symmetric twist}, depends only on a single
integer $k$ coprime with $\hat{L}$, which specifies the flux through
all planes. Indeed, a direct calculation shows that to leading order in 
perturbation theory the volume  dependent term of the plaquette vanishes
exactly, and for other  loops the correction vanishes at large $N$. 
In Ref.~\cite{TEK2} we gave general arguments why this should also happen 
at higher orders  of perturbation theory. 
A detailed analysis of the next-to-leading order is currently under 
way~\cite{MGPAGAMO}, which would allow us to quantify the rate at which 
volume independence  is achieved as a function of $N$ within the perturbative
regime.

The previous paragraphs define our setting. We will be studing SU(N) lattice
gauge theory with Wilson action on a finite box of size $L^4$ with symmetric
twisted boundary conditions and flux $k$. We will also include periodic
boundary conditions, which can be incorporated within the same formalism
but with $k=0$. The only continuous parameter in our study is the lattice 
coupling $b=\beta/(2 N^2)$, the inverse of  `t Hooft coupling $\lambda$ on 
the lattice. Our study includes the  single site ($L=1$) twisted model, 
called Twisted Eguchi-Kawai model (TEK)~\cite{TEK2}, but we also allow for the 
corresponding models with $L=2,4$. For the periodic boundary condition
situation ($k=0$) we have studied various sizes $L=4,8,16,32$. This
allows us to explore the regime studied by Narayanan and
Neuberger~\cite{NN}. These authors proposed that volume independence
holds even for periodic boundary conditions beyond a certain threshold 
$L>L_c(b)$. 

The choice of  $k$ could be crucial, even for values coprime with
$\hat{L}$. Results 
obtained by several authors~\cite{TIMO,TV,Az} showed that at certain
values of the coupling $b$  centre symmetry can break and invalidate the 
EK proof. This problem was analysed by the present authors in
Ref.~\cite{TEK3}, where we concluded that to avoid the problem 
one should take the large $N$ 
limit keeping $k/\hat{L}$ larger than a threshold value of $\sim 0.1$.
Other problems associated with tachyonic instabilities 
require that the limit should be taken keeping $\bar{k}/\hat{L}$ 
large enough too. The integer $\bar{k}$ is defined in terms  of $k$
and $\hat{L}$ as the one satisfying $k\bar{k}=1 \bmod \hat{L}$. These 
requirements match perfectly with corresponding conditions found in a detailed
analysis of the $2+1$-dimensional case~\cite{GPGAO,GPGAO2}. This analysis 
explores the connection between finite volume and finite $N$ effects. 
It seems that the main part of these effects combine into finite volume
effects with an effective volume of $L_{\mathrm{eff}}=L\hat{L}$. 
This result is compatible with volume independence in the large $N$
limit, but goes beyond since it predicts that the main correction 
depends on $L_{\mathrm{eff}}$ (with coefficients that depend on 
$\bar{k}/\hat{L}$). A particular consequence is the possibility of 
replacing  finite-size scaling by finite-$N$ scaling for the model 
with $L=1$, which seems to work remarkably well as seen in a
presentation at this conference~\cite{Keegan}.

\section{Results}
We focused our study upon the behaviour of expectation values of 
small $R\times R$ Wilson loops, from $R=1$ (the plaquette) up to $R=4$. 
These observables are  well measured lattice quantities. 
We explored the dependence of these values upon the parameters of 
the models: the coupling $b$, the matrix rank $N$, the lattice linear
size $L$ and the flux $k$. Here we will restrict to the results obtained 
in the region of relatively strong coupling $0.35 \le b \le 0.385$ which 
corresponds with that from which continuum limit results are 
usually extracted. Our results were obtained with standard Monte Carlo
techniques. Computer resources limits the total number of degrees of
freedom that can be studied. Thus, for $L=32$ we could only go up to
$N=8$, for $L=16$ up to $N=16$ and for $L=1$ up to $N=1369$.

\begin{figure}
\subfloat[]{\includegraphics[width = 0.5\hsize]{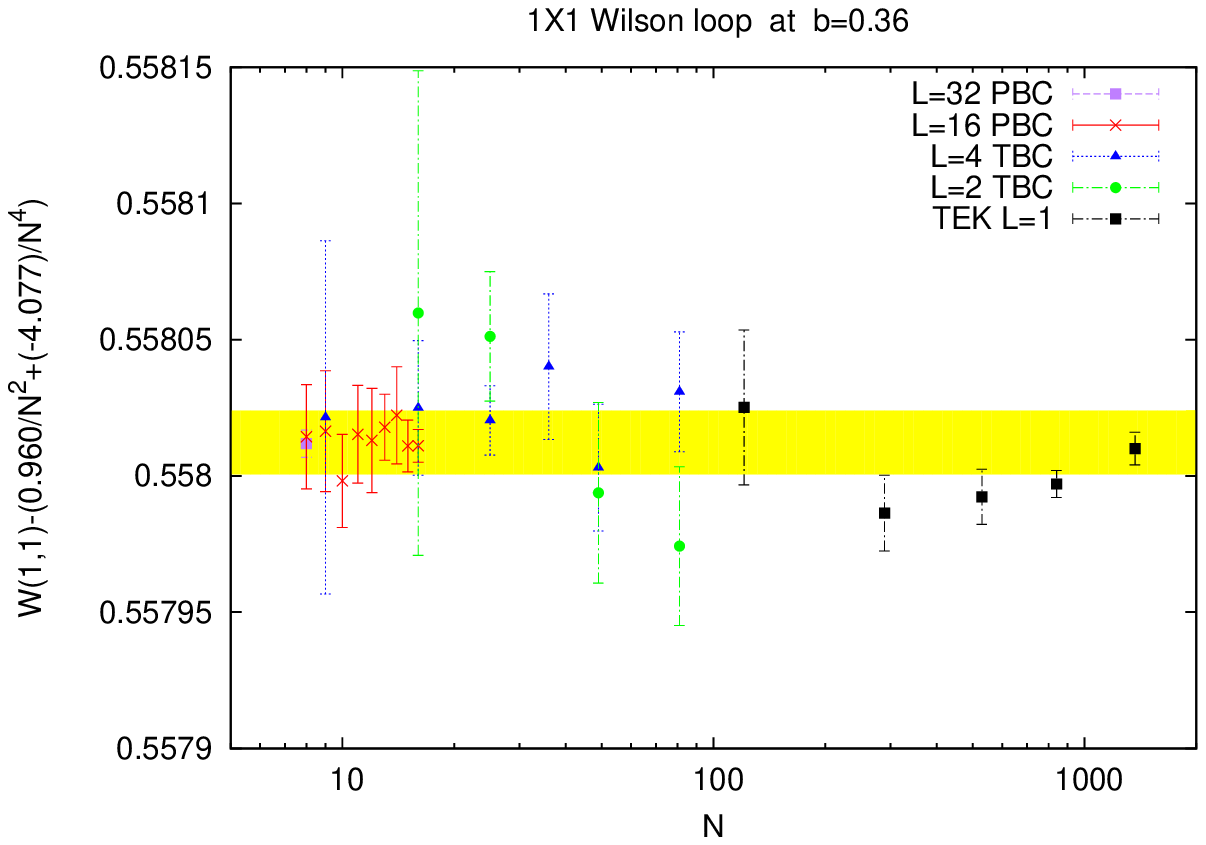}}
\subfloat[]{\includegraphics[width = 0.5\hsize]{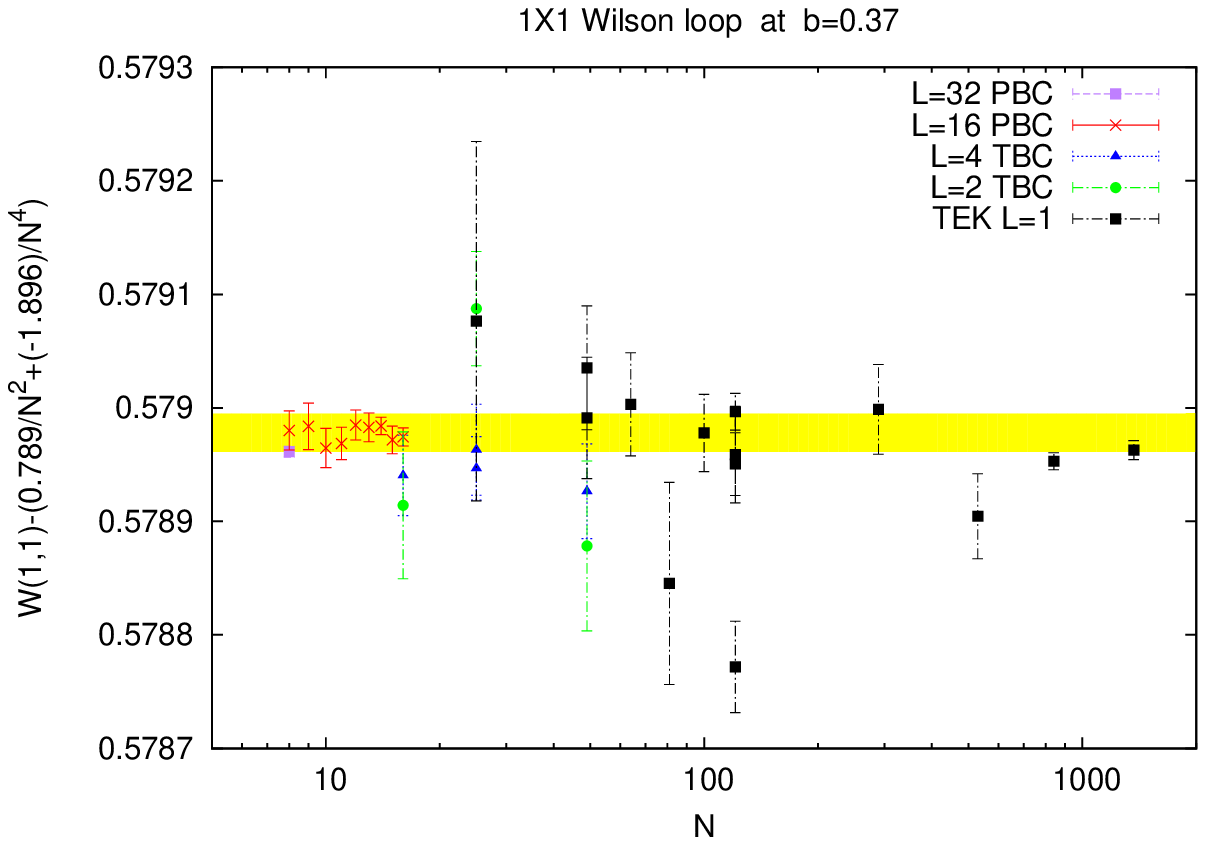}}
\caption{The plaquette expectation value for $b=0.36$ (left) and
$b=0.37$ (right)  after subtracting a term 
$B/N^2+C/N^4$ plotted as a function of $N$. 
The coefficients $B$ and $C$ (appearing in the y-label) are determined
by fitting the $N$ dependence of the periodic
boundary conditions (PBC) $L=16$ points (in red). The one-sigma 
region  in the constant coefficient of the fit is shown as a yellow band.
In addition, we plot 
the $N=8$ $L=32$ plaquette, and the $L=1,2,4$ symmetric twist results;
all with the same subtraction.}
\label{fig1}
\end{figure}

\begin{figure}
\subfloat[]{\includegraphics[width = 0.5\hsize]{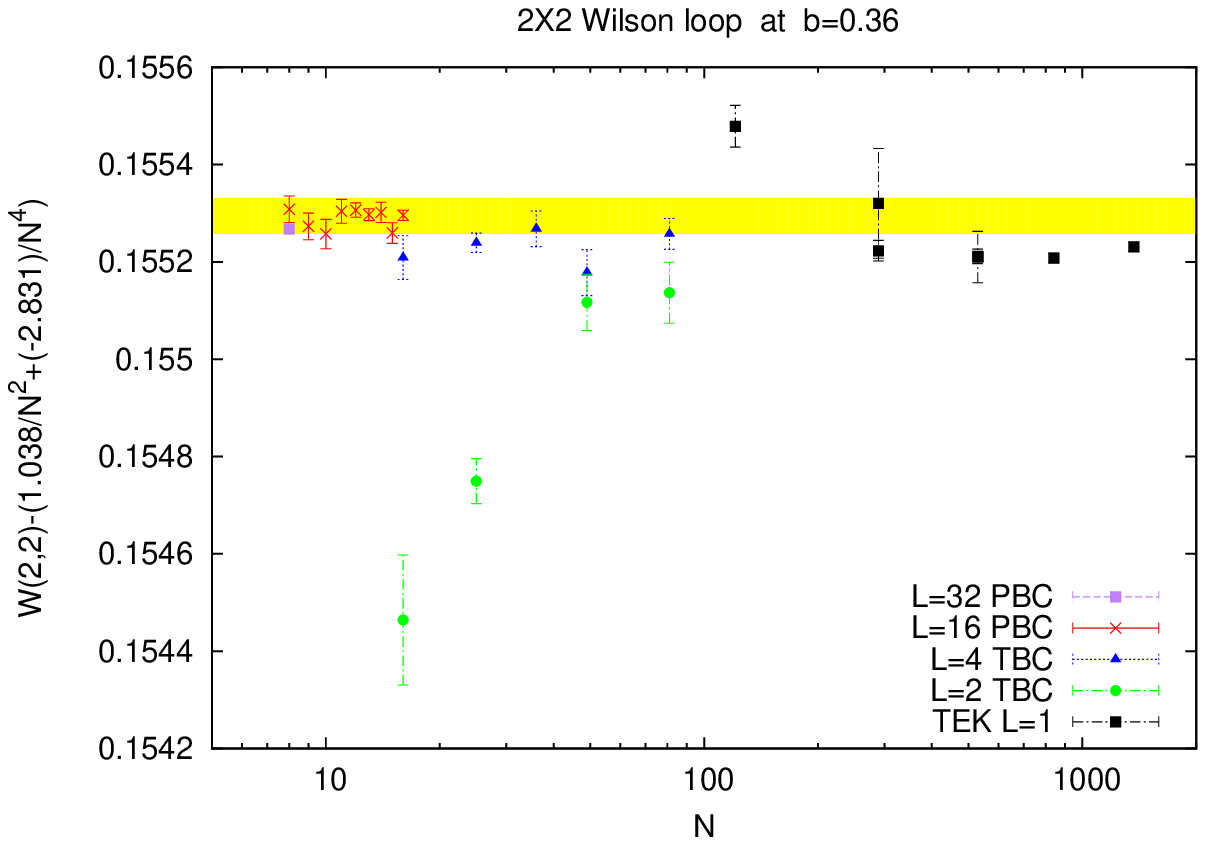}}
\subfloat[]{\includegraphics[width = 0.5\hsize]{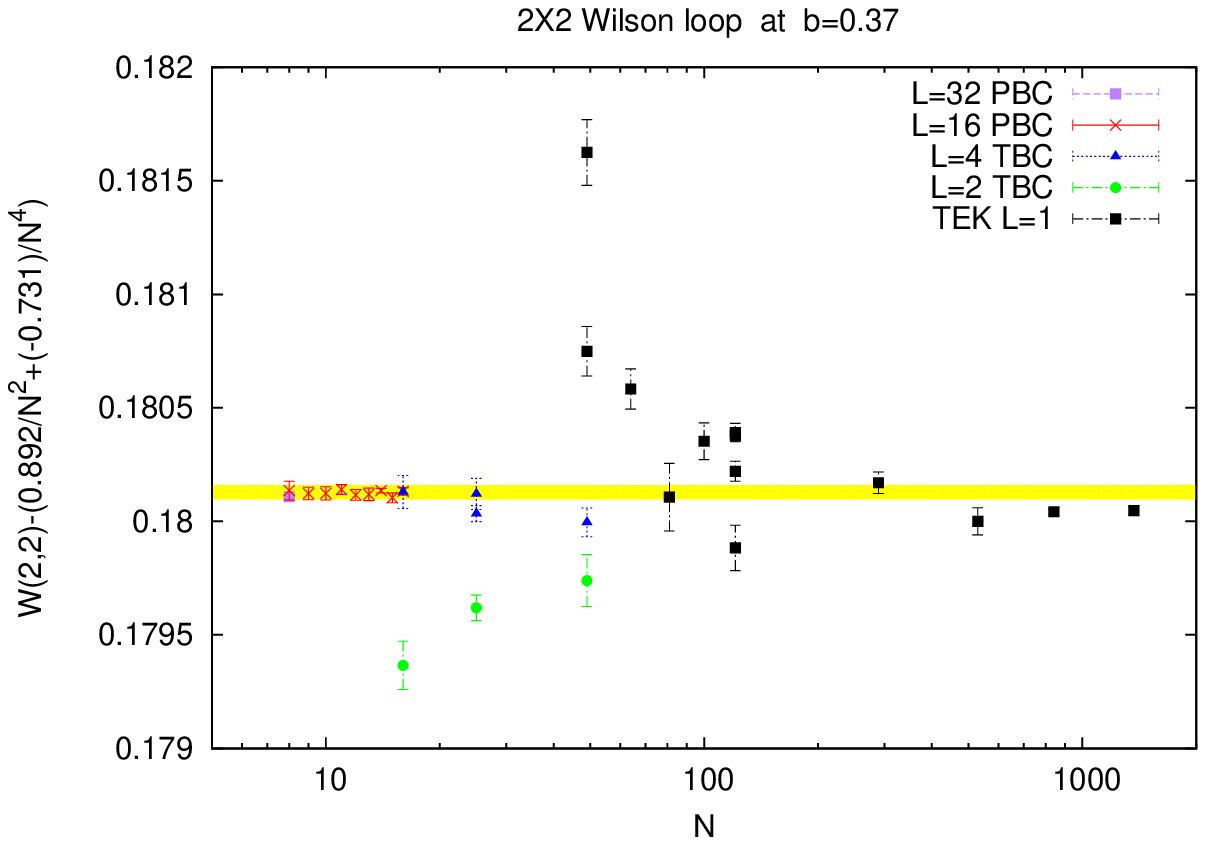}}
\caption{The same as Fig.~1 but for the $2\times 2$ Wilson loop.}
\label{fig2}
\end{figure}

\begin{figure}
\subfloat[]{\includegraphics[width = 0.5\hsize]{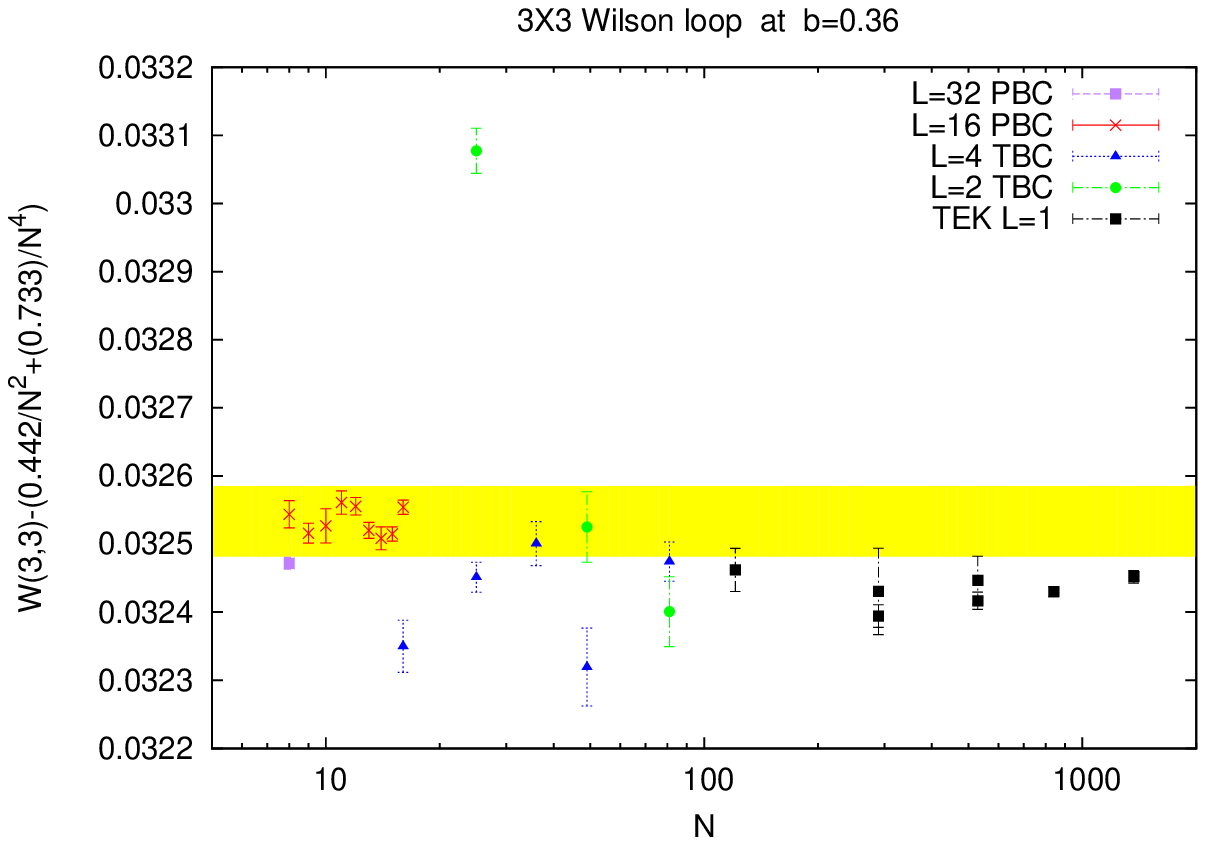}}
\subfloat[]{\includegraphics[width = 0.5\hsize]{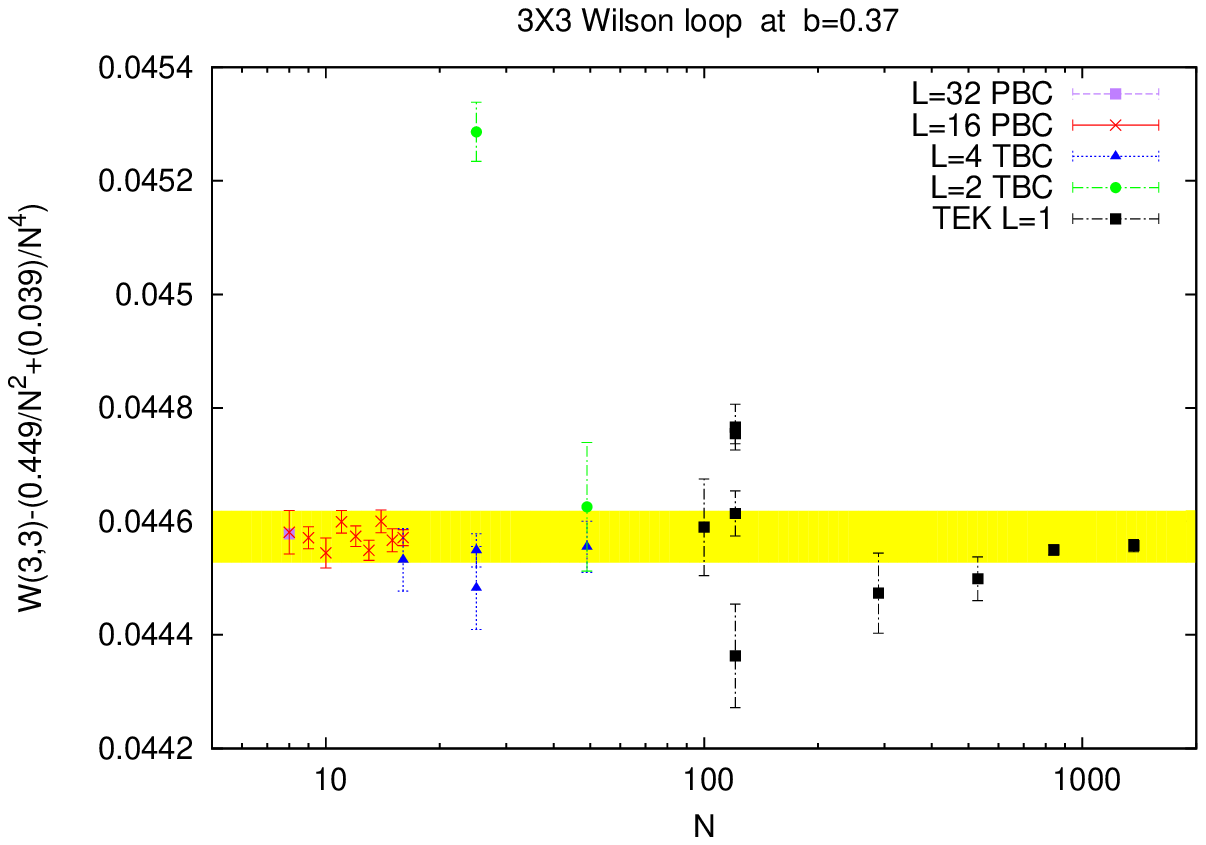}}
\caption{The same as Fig.~1 but for the $3\times 3$ Wilson
loop.}
\label{fig3}
\end{figure}

\begin{figure}
\subfloat[]{\includegraphics[width = 0.5\hsize]{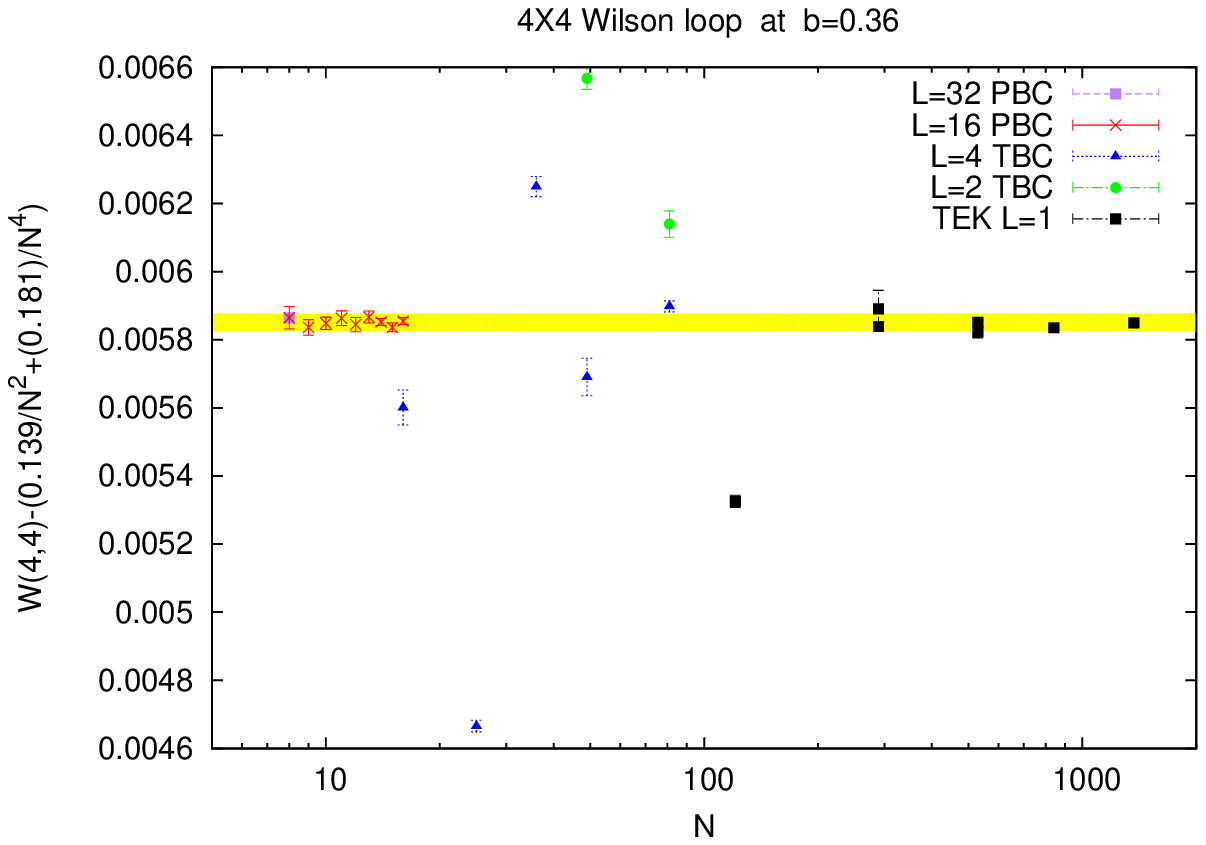}}
\subfloat[]{\includegraphics[width = 0.5\hsize]{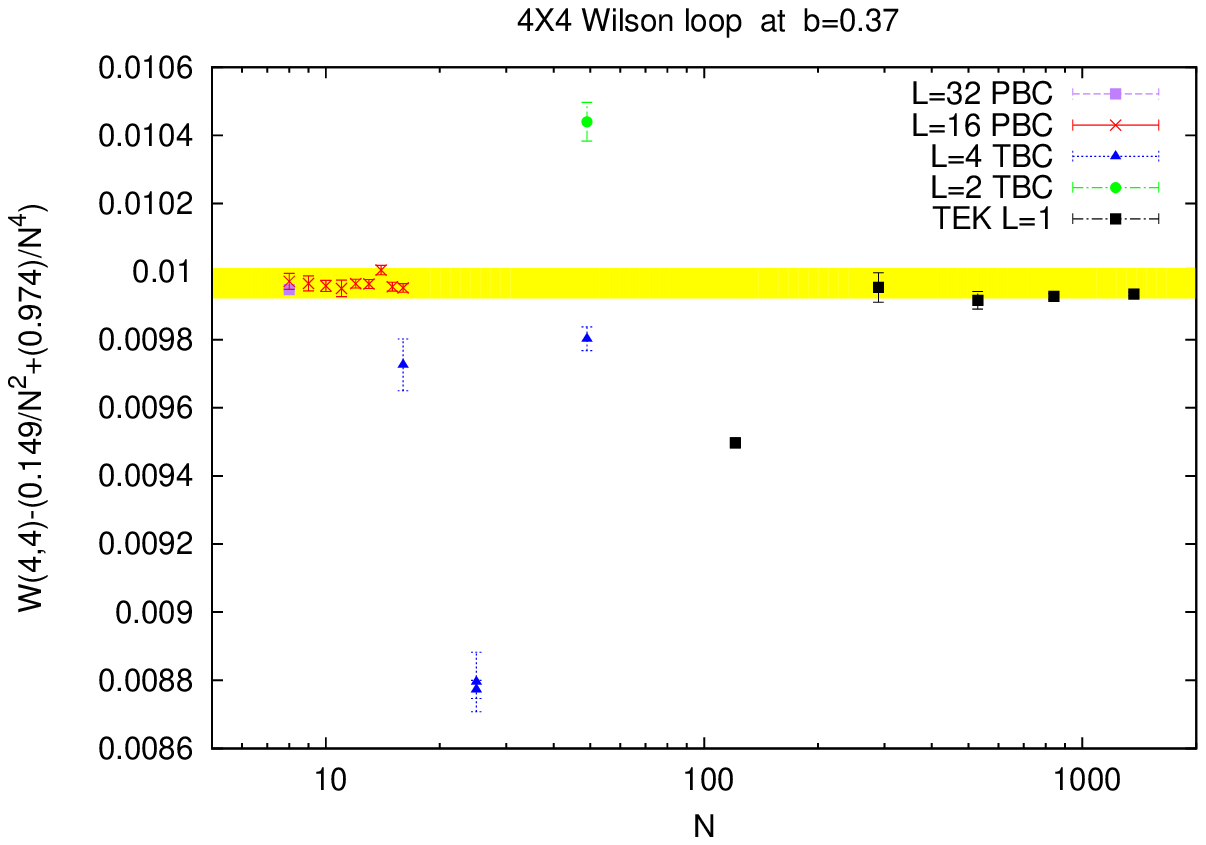}}
\caption{The same as Fig.~1 but for the $4\times 4$ Wilson
loop.}
\label{fig4}
\end{figure}

A  systematic analysis was performed at two values of the coupling
$b=0.36$ and $b=0.37$. We started by studying the system with periodic
boundary conditions ($k=0$). At fixed $N$, volume dependence is clearly observed in the 
expectation values for $L<16$. The correction turns out to be positive
for all our observables. On the other hand, the results at $L=16$ 
are compatible within errors with those at $L=32$ wherever the two
were available. Thus, we can take the $L=16$ results to be a good
approximation to those at infinite volume. 
At $L=16$ we performed 
simulations at all values of $N$ in the interval $[8,16]$. The $N$
dependence of the results is sizable at the scale of the errors. 
Thus, to obtain good estimates of the value of these observables 
at $N=\infty$, we should extrapolate our results. Good fits are
obtained in all cases with a quadratic polynomial in $1/N^2$: 
$A+B/N^2+C/N^4$. The  correction is dominated by the $1/N^2$ correction with a 
coefficient $B$ which is positive and  of order 1 (dependent on $b$ and $R$). 
For display  purposes we subtracted $B/N^2+C/N^4$ from the measured
values and displayed them as the red points in Figs.~\ref{fig1}-\ref{fig4}.
The yellow band gives the value of $A$ within one sigma obtained from the
fit. This value amounts to the determination of the corresponding
observable at $N=\infty$. In the same plot we also display the
measurement obtained for $L=32$ and $N=8$. 

As mentioned earlier the expectation values obtained for $L=8$ are
significantly different than those of $L=16$ however a similar fit for
these  data allows to extrapolate to $N=\infty$. Curiously the fit
obtained for $b=0.36$ gives a consistent extrapolation for all loops
except $R=4$. On the contrary the extrapolation for $b=0.37$ is
inconsistent with that of $L=16$. Repeating the process for $L=4$ (for
which we could explore larger values of $N$) we
found clearly inconsistent extrapolations in all cases. These results
are in agreement with the conclusions of Ref.~\cite{NN} since
$L_c(0.37)> L_c(0.36)\sim 8$.

We also studied the results obtained with the symmetric twist and
$L=1,2,4$. The raw values are displayed in the same
figs.~~\ref{fig1}-\ref{fig4} after subtracting  $B/N^2+C/N^4$
with $B$ and $C$ obtained from the $L=16$ periodic data (no new fit).
The results for $N>300$ obtained with the single site TEK model 
are essentially unaffected by the subtraction. These results include 
the case of $N=529$ ($\hat{L}=23$, $k=7$, $\bar{k}=10$), $N=841$ 
($\hat{L}=29$, $k=9$, $\bar{k}=13$), and $N=1369$ ($\hat{L}=37$, 
$k=11$, $\bar{k}=10$). The results are compatible with each
other and with the  extrapolation of the $L=16$ results. This is quite 
remarkable given the  wide range of $N$ values covered, the small errors
of the data  and the large relative finite $N$ corrections to the $L=16$
results. This provides a strong confirmation of volume independence 
in the extreme case, since we compare results in the infinite volume
limit with those of $L=1$. Obviously twisted boundary conditions are
essential for this result. Notice that the values of the flux $k$ lie 
within the safe region. In some cases (for example the $3\times 3$ 
loop at $b=0.36$) the TEK data lies below the yellow band, although 
these  deviations are smaller than 2 sigma. In any case, we believe 
that the TEK value is a more reliable estimate of the large $N$ result 
than the extrapolated result, given the characteristic systematic
errors of extrapolations.

In Figs.~\ref{fig1}-\ref{fig4} we also plot the results obtained for 
smaller values of $N$ at $L=1,2,4$ and symmetric twisted boundary
conditions applying the same subtraction as before. We plotted all 
the available results which matched the 
criteria $\frac{\bar{k}}{\hat{L}}>0.25$ and
$L_{\mathrm{eff}}=\hat{L}L>3 R$. Extending the region, the deviations
grow and the scale of the plot has to be reduced accordingly. Two
comments are in order. The first is that there is no reason why the 
$1/N^2$ corrections of the twisted data should coincide with those 
of the $L=16$ periodic data. The second is that these corrections
might depend on the value of $k$. However, the fact that all the results 
fall within the scale of the plot implies that the corrections are not 
terribly different from the periodic ones as long as we stay within
the safe $k$ and $\bar{k}$ range. This is particularly remarkable for
the plaquette  for which almost all the data fit nicely within the 
extrapolated band.  We do not have enough information for a more 
systematic study of the $k$ dependence of the corrections, but that
is  not necessary for the specific goals of this work.

We have performed less extensive tests at other values of $b$ and found
results consistent with those of $b=0.36$ and $0.37$. Furthermore, we have  
been able to find interpolating formulas that allow us to use all the
available data both for large sizes and periodic boundary conditions 
and for the TEK model. All the data are consistent with the latter
matching the infinite $N$ limit of the former. The full results can be 
consulted in Ref.~\cite{GAO}

\section{Conclusions}
We have analyzed Wilson loop expectation values for various values of 
the lattice size $L$ and $N$ for both periodic and symmetric twisted
boundary conditions. The results obtained by large $N$ extrapolation 
and large volumes match within the tiny errors of the data with those 
obtained with the $L=1$ TEK model. This provides a strong direct
evidence supporting volume independence at large $N$.

\section*{Acknowledgments}
A.G-A acknowledges financial support from the grants FPA2012-31686
and FPA2012-31880, the MINECO Centro de Excelencia Severo Ochoa Program SEV-
2012-0249, the Comunidad AutLonoma de Madrid HEPHACOS S2009/ESP-1473, and
the EU PITN-GA-2009-238353 (STRONG net). He participates in the Consolider-
Ingenio 2010 CPAN (CSD2007-00042). M. O. is supported by 
the Japanese MEXT grant No 26400249. 

Calculations have been done on  Hitachi SR16000 supercomputer 
both at High Energy  Accelerator Research  Organization(KEK) and YITP in Kyoto University.
 Work at  KEK is supported by the Large Scale Simulation Program No.13/14-02. 
Calculations have also been done at INSAM clusters at Hiroshima
University and HPC-clusters and servers at IFT.


\vspace{-0.2cm}
  
\begin{thebibliography}{99}

\bibitem{EK}
  T.~Eguchi and H.~Kawai, 
  {\it Reduction of dynamical degrees of freedom in the large N gauge theory},
  Phys.\ Rev.\ Lett.\  {\bf 48} (1982) 1063.

 \bibitem{GAO}
  A.~Gonzalez-Arroyo and M.~Okawa,
  {\it Testing volume independence of SU(N) pure gauge theories at large N},
  [arXiv:1410.6405 [hep-lat]]. 

\bibitem{bhanot}
  G.~Bhanot, U.~M.~Heller and H.~Neuberger,
  {\it The quenched Eguchi-Kawai model},
  Phys.\ Lett.\  B {\bf 113} (1982) 47.


\bibitem{TEK1}
  A.~Gonzalez-Arroyo and M.~Okawa,
    {\it A twisted model for large N lattice gauge theory},
      Phys.\ Lett.\  B {\bf 120} (1983) 174.
      
      


\bibitem{TEK2}
  A.~Gonzalez-Arroyo and M.~Okawa,
  {\it Twisted Eguchi-Kawai model: A reduced model 
  for large N lattice gauge theory},
  Phys.\ Rev.\  D {\bf 27} (1983) 2397.



\bibitem{MGPAGAMO}
M.~Garcia~Perez, A.~Gonzalez-Arroyo and M.~Okawa,
{\it Perturbative contributions to Wilson loops in twisted lattice boxes and reduced models},	  
	In preparation. 

\bibitem{NN}
  R.~Narayanan and H.~Neuberger,
  {\it Large N reduction in continuum},
  Phys.\ Rev.\ Lett.\  {\bf 91} (2003) 081601
  [arXiv:hep-lat/0303023];
  J.~Kiskis, R.~Narayanan and H.~Neuberger,
  {\it Does the crossover from perturbative to nonperturbative physics in
   QCD become a phase transition at infinite N?},
    Phys.\ Lett.\ B {\bf 574} (2003) 65
      [hep-lat/0308033].

\bibitem{TIMO}
  T.~Ishikawa and M.~Okawa, 
  {\it $Z^D_N$ symmetry breaking on the numerical simulation of twisted
  Eguchi-Kawai model}, 
  talk given at the Annual Meeting of the Physical Society of Japan,
  March 28-31, Sendai, Japan (2003).

\bibitem{TV}
  M.~Teper and H.~Vairinhos,
  {\it Symmetry breaking In twisted Eguchi-Kawai models},
  Phys.\ Lett.\  B {\bf 652} (2007) 359
  [arXiv:hep-th/0612097].

\bibitem{Az}
  T.~Azeyanagi, M.~Hanada, T.~Hirata and T.~Ishikawa,
  {\it Phase structure of twisted Eguchi-Kawai model},
  JHEP {\bf 0801} (2008) 025
  [arXiv:0711.1925 [hep-lat]]. 
  


\bibitem{TEK3}
  A.~Gonzalez-Arroyo and M.~Okawa, 
  {\it Large N reduction with the Twisted Eguchi-Kawai model},
  JHEP {\bf 1007} (2010) 043
  [arXiv:1005.1981 [hep-th]].  
  
  
\bibitem{GPGAO}
  M.~Garcia~Perez, A.~Gonzalez-Arroyo and M.~Okawa,
  {\it Spatial volume dependence for 2+1 dimensional SU(N) Yang-Mills theory},
  JHEP {\bf 1309} (2013) 003
  [arXiv:1307.5254 [hep-lat]].
  

\bibitem{GPGAO2}
  M.~Garcia~Perez, A.~Gonzalez-Arroyo and M.~Okawa,
  {\it Volume independence for Yang-Mills fields on the twisted torus},  
  International Journal of Modern Physics A Vol. 29, No. 25 (2014) 1445001, 
  [arXiv:1406.5655 [hep-th]].	  

\bibitem{Keegan}
  M.~Garcia~Perez, A.~Gonzalez-Arroyo. L.~Keegan and M.~Okawa,
  {\it TEK twisted gradient flow running coupling},  
  PoS {\bf LAT2014} (2014) 300 [arXiv:1411.0258 [hep-lat]].




  
\end{thebibliography}
\end{document}